\documentclass[11pt]{article}

\usepackage{amssymb,amsmath,epsfig}

\usepackage{amsmath}
\usepackage{amssymb}
\usepackage{amsthm}
\usepackage{url}
\usepackage{graphicx} 
\usepackage{rotating}

\newcommand\x{{\bf x}}
\newcommand\y{{\bf y}}

\newcommand\bfa{{\bf a}}
\newcommand\bfb{{\bf b}}
\newcommand\bfc{{\bf c}}

\newcommand\bfu{{\bf u}}
\newcommand\bfv{{\bf v}}
\newcommand\bfw{{\bf w}}
\newcommand\zero{{\bf 0}}
\newcommand\cc{{\mathbb C}}
\newcommand\zz{{\mathbb Z}}
\newcommand\qq{{\mathbb Q}}

\newtheorem{theorem}{Theorem}[section]

\newtheorem{proposition}[theorem]{Proposition}

\theoremstyle{definition}

\begin{document}
\title{Computing Puiseux Series for Algebraic Surfaces\thanks{This
material is based upon work supported by the National Science Foundation
under Grant No.\ 1115777.} }
\author{Danko Adrovic and Jan Verschelde\\
Department of Mathematics, Statistics, and Computer Science \\
University of Illinois at Chicago \\
851 South Morgan (M/C 249) \\
Chicago, IL 60607-7045, USA \\
  \texttt{adrovic@math.uic.edu, jan@math.uic.edu}\\
  \texttt{\url{www.math.uic.edu/~adrovic}, \url{www.math.uic.edu/~jan}}}

\date{30 April 2012}

\maketitle

\begin{abstract}
In this paper we outline an algorithmic approach to compute Puiseux
series expansions for algebraic surfaces.   
The series expansions originate at the intersection of the surface 
with as many coordinate planes as the dimension of the surface.  
Our approach starts with a polyhedral method to compute cones of normal 
vectors to the Newton polytopes of the given polynomial system 
that defines the surface.  
If as many vectors in the cone as the dimension of the surface define an 
initial form system that has isolated solutions, then those vectors are 
potential tropisms for the initial term of the Puiseux series expansion.
Our preliminary methods produce exact representations for solution sets
of the cyclic $n$-roots problem, for $n = m^2$, corresponding to a result
of Backelin.

\medskip

\noindent {\bf Keywords.}
algebraic surface,
binomial system,
cyclic $n$-roots problem,
initial form,
Newton polytope, 
orbit,
permutation symmetry,
polyhedral method,
Puiseux series, 
sparse polynomial system, 
tropism,
unimodular transformation.

\end{abstract}

\section{Introduction}

We presented polyhedral algorithms to develop Puiseux expansions, 
for plane curves in~\cite{AV11a} and for space curves in~\cite{AV11b},
based on ideas described in~\cite{Ver09b}.
In this paper we explain a polyhedral approach to compute series
developments for algebraic sets.  Although we use the numerical
solver of PHCpack~\cite{Ver99}, one may use any solver for the
leading coefficients of the series and obtain a purely symbolic method.
We implemented our methods using Sage~\cite{Sage}.

We could reduce the treatment of algebraic sets to the curve case
by adding sufficiently many hyperplanes in general position to cut out
a curve on the set.  This approach~\cite{SVW05} is not flexible enough
to exploit permutation symmetry as the added general hyperplanes
ignore the symmetric structure of the polynomial system.

Although presently we do not have a fully automatic implementation
suitable for benchmarking on a large class of polynomial systems,
we have obtained promising results on the cyclic $n$-root systems:

\begin{equation} \label{eqcyclicsys}
   \begin{cases}
   x_{0}+x_{1}+ \cdots +x_{n-1}=0 \\
   x_{0}x_{1}+x_{1}x_{2}+ \dots +x_{n-2}x_{n-1}+x_{n-1}x_{0}=0 \\
   i = 3, 4, \ldots, n-1: 
    \displaystyle\sum_{j=0}^{n-1} ~ \prod_{k=j}^{j+i-1}
    x_{k~{\rm mod}~n}=0 \\
   x_{0}x_{1}x_{2} \cdots x_{n-1} - 1 = 0. \\
\end{cases}
\end{equation}

The cyclic $n$-roots system is a standard benchmark problem
in computer algebra, relevant to operator algebras.  
We refer to~\cite{Szo11} for recent advances in
the classification of complex Hadamard matrices.
In~\cite{Fau01}, the close relationship of~(\ref{eqcyclicsys}) with
some systems occurring in optimal design of filter banks is stressed.
The numerical factorization of the two dimensional surface of
cyclic 9-roots into 6 irreducible cubics was reported in~\cite{SVW01b}.
Recent results for the cyclic 12-roots problem can be found in~\cite{Sab11}.

Surprisingly, while looking to develop Puiseux series for
algebraic sets, for cyclic 9-roots we found exact results:
the first term of the series satisfies the entire polynomial system.
These exact result correspond to known (see e.g.~\cite{Bac89} 
or~\cite{Fau01}) configurations of cyclic $n$-roots.

The type of polynomial systems targeted by the polyhedral approach
are sparse polynomial systems.  We introduce our approach in the next
section with a very particular sparse class of systems.
We use unimodular transformations to work with points at infinity.
The second section ends with a general approach to solve a binomial system.

To find the initial coefficients in the Puiseux series we look for
initial form systems, systems that have fewer monomials than the
original systems and that are supported on faces of the Newton polytopes.
Faces of the Newton polytopes that define the initial forms
are determined by their inner normals.  Those inner normals that define
the initial form systems are the leading powers (called tropisms)
of generalized Puiseux series.  The leading coefficients of the series 
vanish at the initial form systems.

In the third section we define initial form systems, give an
illustrative example, and describe the degeneration of a $d$-dimensional
algebraic set along a path towards the intersection with the first $d$
coordinate planes.  Polyhedral methods give us cones of pretropisms
and initial form systems that may lead to initial coefficients of
Puiseux series.  We end this paper giving an exact description of
positive dimensional sets of cyclic $n$-roots.

\noindent {\bf Related work.}  A geometric resolution of a polynomial
system uses a parameterization of the coordinates~\cite{GLS01} for
global version of Newton's iterator~\cite{CPHM99}.
Our algorithms arose from an understanding of~\cite[Theorem~B]{Ber75}
and are inspired by tropical methods~\cite{BJSST07} and 
in particular by the constructive proof of the fundamental theorem
of tropical algebraic geometry~\cite{JMM08}.  
Puiseux series occur perhaps most often in the resolution of
singularities, \cite{AIL11} describes an extension of Newton's method 
using the notion of tropical variety.
Software related to~\cite{JMM08} is Gfan~\cite{Jen08} and
the Singular library {\tt tropical.lib}~\cite{JMM07}.

Connections with Gr\"obner bases are described in~\cite{Stu96}.
Polyhedral and tropical methods for finiteness proofs in celestial 
mechanics are explained in~\cite{HM06} and~\cite{JH11}.
Truncations of two dimensional varieties are studied in~\cite{Kaz99}.
The unimodular coordinate transformations
are related to power transformations in~\cite{Bru00}.
A Newton-Puiseux algorithm for polynomials in several variables is
described in~\cite{BR03}.  
In~\cite{McD02}, fractional power series solutions are developed
for generic systems.

\noindent {\bf Acknowledgements.}  We thank Marc Culler for mentioning
the Smith normal form for unimodular transformations.
We appreciate the comments of the reviewers.

\section{Binomial Systems}

We aim to solve {\em sparse} polynomial systems, systems of polynomials
with relatively few monomials appearing with nonzero coefficient.
The sparsest polynomial systems which admit solutions with nonzero
values for all coordinates consist of {\em exactly two} monomials in every
equation and we call such systems {\em binomial} systems.
See e.g.:~\cite{CD07} and~\cite{DMM10} for more on binomial ideals.

To represent a $d$-dimensional solution set~$S$ intersecting the first
$d$ coordinate planes in as many regular isolated points as the degree
of~$S$, the first $d$ variables can serve as independent parameters.
The parameterizations that are of interest to us start with the generators
of cones of normal vectors defining initial forms of polynomial systems.

\subsection{An Example}

Consider for example
\begin{equation} \label{eqbinsys}
   \left\{
      \begin{array}{l}
         x_0^2 x_1 x_2^4 x_3^3 - 1 = 0 \\
         x_0 x_1 x_2 x_3 - 1 = 0.
      \end{array}
   \right.
\end{equation}
We write the exponent vectors in the matrix
\begin{equation} \label{eqexpmatrix}
   A = \left[
      \begin{array}{cccc}
         2 & 1 & 4 & 3 \\
         1 & 1 & 1 & 1 
      \end{array}
   \right]
\end{equation}
and we look for a basis of the null space of~$A$.
Two linearly independent vectors that satisfy~$A \x = \zero$
are for example $\bfu = (-3,2,1,0)$ and~$\bfv = (-2,1,0,1)$.
Placing $\bfu$ and~$\bfv$ in the columns of a matrix~$M$
leads to a coordinate transformation:
\begin{equation} \label{eqcoordtrans}
   M = \left[
     \begin{array}{rrrr}
        -3 & -2 & 1 & 0 \\
         2 &  1 & 0 & 1 \\
         1 &  0 & 0 & 0 \\
         0 &  1 & 0 & 0 \\
     \end{array}
   \right]
   \quad
   \left\{
      \begin{array}{l}
         x_0 = y_0^{-3} y_1^{-2} y_2 \\
         x_1 = y_0^2 y_1 y_3 \\
         x_2 = y_0 \\
         x_3 = y_1.
      \end{array}
   \right.
\end{equation}
The coordinate transformation $\x = \y^M$ eliminates $y_0$ and~$y_1$
--- because $\bfu$ and $\bfv$ are in the null space of~$A$ ---
as substituting the coordinates corresponds to computing $A \bfu$
and $A \bfv$, reducing the given system to
\begin{equation} \label{eqredsys}
  \left\{
     \begin{array}{l}
        y_2^2 y_3 - 1 = 0 \\
        y_2 y_3 - 1 = 0.
     \end{array}
  \right.
\end{equation}
Solving the reduced system in~(\ref{eqredsys})
gives values for $y_2$ and~$y_3$ which after substitution
in the coordinate transformation in~(\ref{eqcoordtrans})
yields an explicit solution for the original system in~(\ref{eqbinsys})
with $y_0$ and~$y_1$ as parameters.

\subsection{Unimodular Transformations}

In the previous section we constructed in~(\ref{eqcoordtrans})
a unimodular coordinate transformation $\x = \y^M$, where~$\det(M) = \pm 1$.  
In the new $\y$ coordinates all points that make the same inner product
of the $i$th row of the given exponent matrix~$A$ will have the same
value for~$y_i$.  

The null space of the matrix~$A$ is stored in the rows 
of the matrix~$B$: $A B^T = \zero$.
The Smith normal form of~$B$ consists of the triplet
$(U,S,V)$, where $U$ and $V$ are unimodular matrices
($\det(U) = \pm 1$ and $\det(V) = \pm 1$), 
and the only nonzero elements of~$S$ are on the diagonal: $U B V = S$.

If $U$ equals the identity matrix, then $U B V = S$
implies $B = S V^{-1}$.   This means that for any~$\x$,
the outcome of $B \x$ is the same as $S V^{-1} \x$.
If moreover~$S$ contains the identity matrix, then $V^{-1}$
defines the unimodular transformation~$M$.
The next examples illustrates the case of general~$U$
but where $S$ contains the identity matrix.

For the matrix~$A$ in~(\ref{eqexpmatrix}), the matrix~$B$
has in its two rows the vectors~$\bfu$ and~$\bfv$
so that $A B^T = \zero$:
\begin{equation}
   B = \left[
      \begin{array}{rrrr}
         -3 & 2 & 1 & 0 \\
         -2 & 1 & 0 & 1
      \end{array}
   \right].
\end{equation}
The computation of the Smith normal form of~$B$ with GAP~\cite{GAP}
(from the console in Sage~\cite{Sage}) gives
\begin{equation}
   U = \left[
      \begin{array}{rr}
         1 & -2 \\
         2 & -3
      \end{array}
   \right], \quad
   S = \left[
      \begin{array}{rrrr}
         1 & 0 & 0 & 0 \\
         0 & 1 & 0 & 0
      \end{array}
   \right],
\end{equation}
and
\begin{equation}
   V = \left[
      \begin{array}{rrrr}
         1 & 0 & 1 & -2 \\
         0 & 1 & 2 & -3 \\
         0 & 0 & 1 & 0 \\
         0 & 0 & 0 & 1
      \end{array}
   \right].
\end{equation}
We use the inverses $U^{-1}$ and $V^{-1}$ to construct
a unimodular transformation extending $U^{-1}$ with
the identity matrix, as follows:
\begin{equation}
   \left[
      \begin{array}{rrrr}
         -3 & 2 & 0 & 0 \\
         -2 & 1 & 0 & 0 \\
          0 & 0 & 1 & 0 \\
          0 & 0 & 0 & 1 
      \end{array}
   \right]
   \left[
      \begin{array}{rrrr}
         1 & 0 & 1 & -2 \\
         0 & 1 & 2 & -3 \\
         0 & 0 & 1 & 0 \\
         0 & 0 & 0 & 1
      \end{array}
   \right]
\end{equation}
and this product gives the transpose of~$M$,
the matrix in the unimodular transformation of~(\ref{eqcoordtrans}).
This examples illustrates the case when $U$ is not the identity matrix
and where we may ignore $S$ as its diagonal elements are all equal to one.

We point out that the vectors in the null space of the exponent
matrix~$A$ as in~(\ref{eqexpmatrix}) are typically normalized
so that the greatest common divisors of the components of the vectors
equals one.  We may change coordinates so that the first vector in
the null space has only its first coordinate different from zero,
the second vector in the null space can have nonzero entries only
in the first two coordinates, etc.  

Although we prefer to represent the solution set using only integer
exponents for the parameters, this is not always possible, consider
for example
\begin{equation}
   B = \left[
      \begin{array}{rrrr}
         2 &  6 & 17 & 9 \\
         4 & 14 & 13 & 3
      \end{array}
   \right].
\end{equation}
The divisors for the two rows of~$B$ (and the denominators of
the exponents of the parameters) are obtained via the Hermite
normal form of~$B$: $U B = H$, where $U$ is a square unimodular matrix
and~$H$ an upper triangular matrix.  We assume that $B$ is full rank
and that the columns have been permuted so $H$ has only nonzero
elements on its diagonal.  Let $D$ be a diagonal matrix of
the same dimensions as~$U$ which takes its elements from the corresponding
diagonal elements of the matrix~$H$.  Then the coordinate transformation
is defined by
\begin{equation} \label{eqMhermite}
   M = \left[
     \begin{array}{cc}
        \multicolumn{2}{c}{D^{-1} B} \\
         \zero & I
     \end{array}
   \right].
\end{equation}
where $I$ is the identity matrix.  To show that the determinant of~$M$
equals $\pm 1$, consider the extended unimodular matrix
\begin{equation}
   \widehat{U} =
   \left[
      \begin{array}{cc}
         U & \zero \\
         \zero & I
      \end{array}
   \right].
\end{equation}
Because $U$ is unimodular, $\det(\widehat{U}) = \pm 1$ and
$\det(\widehat{U} M) = \pm \det(M)$.
We have $\det(\widehat{U} M) = \pm 1$, because $\widehat{U} M$
is an upper triangular matrix with $\pm 1$ on its diagonal as
a result of the multiplication by~$D^{-1}$.

Note that the rational exponents will appear only in the powers
of the parameters as performing the coordinate transformation
$\x = \y^M$ on the system $\x^A - \bfc = \zero$ eliminates the
first $d$ variables of the $d$-dimensional solution set.

\subsection{Solving Binomial Systems}

We denote a binomial system by $\x^A - \bfc = \zero$,
where $A \in \zz^{k \times n}$ and $\bfc = (c_0,c_1,\ldots,c_{k-1})^T$
with $c_i \not= 0$ for all $i = 0,1,\ldots,k-1$.  If the rank of~$A$
equals $k$, then $k$ is the codimension of the solution set.
Given the tuple $(A,\bfc)$, the solution set of~$\x^A - \bfc = \zero$
is described by a unimodular transformation~$M$ and a set of
values for the last $n-k$ variables.  

In the sketch of the solution method below we assume that~$A$
has rank~$k$, otherwise $\x^A - \bfc = \zero$ has no $(n-k)$-dimensional
solution set for general values of~$\bfc$.  The steps are as follows:

\begin{enumerate}
\item Compute the null space $B$ of~$A$, $d = n - k$.
\item Compute the Smith normal form $(U,S,V)$ of~$B$.
\item Depending on $U$ and $S$ do one of the following:
\begin{itemize}
  \item If $U$ is the identity matrix, then $M = V^{-1}$
        and the first $d$ variables have positive denominators
        in their powers when not all elements on the diagonal of $S$ 
        are equal to one.
  \item If $U$ is not the identity matrix and
        if all elements on the diagonal of~$S$ are one,
        then extend $U^{-1}$ with an identity matrix to obtain
        an $n$-by-$n$ matrix~$E$ that has $U^{-1}$ in its first
        $d$ rows and columns.  Then, $M = E V^{-1}$.
  \item In all other cases, define~$M$ as in~(\ref{eqMhermite}).
\end{itemize}
\item After the coordinate transformation~$\x = \y^M$,
      compute the leading coefficients solving a binomial system 
      of~$k$ equations in~$k$ unknowns.
      Return $M$ and the corresponding solutions of the binomial system.
\end{enumerate}

The solution procedure for binomial systems outlined above returns
a representation with $d$ parameters for the $d$-dimensional solution set
which can geometrically interpreted as follows.  
For zero values of the parameters, we obtain the points of the solution set
intersected with the first $d$ coordinate hyperplanes.
For nonzero values of the parameters, the powers of the parameters
correspond to a choice of the basis for the null space of the exponent
matrix of the binomial system.

\section{Sparse Polynomial Systems}

To look for $d$-dimensional components of sparse polynomial systems,
we investigate solutions of initial forms defined by cones of
normal vectors.  In order for the initial form systems to have solutions
with all coordinates different from zero, they need to be at least
binomial systems.

Although not all (and perhaps only few) initial form systems are
binomial, the unimodular transformations explained in \S2.2 
are applied on a matrix of pretropisms.

\subsection{Initial Forms}

A polynomial $f$ in $n$ variables $\x = (x_0,x_1,\ldots,x_{n-1})$ is
denoted as
\begin{equation}
   f(\x) = \sum_{\bfa \in A} c_\bfa \x^\bfa, \quad
   c_\bfa \in \cc \setminus \{ 0 \},
\end{equation}
$x^\bfa = x_0^{a_0} x_1^{a_1} \cdots x_{n-1}^{a_{n-1}}$, 
where $A$ is the set of all exponents of monomials with nonzero coefficient.
The set $A$ is the {\em support} of~$f$ and 
the convex hull of~$A$ is the {\em Newton polytope} $P$ of~$f$.
Any nonzero vector $\bfv$ defines a face of~$P$, spanned by
\begin{equation}
   {\rm in}_\bfv(A) = \{ \ \bfb \in A \ | \
   \langle \bfb , \bfv \rangle 
   = \min_{\bfa \in A} \langle \bfa, \bfv \rangle \ \},
\end{equation}
where $\langle \cdot, \cdot \rangle$ denotes the usual inner product
of two vectors.  We use the notation ${\rm in}_\bfv(A)$ because a face
of a support set defines an {\em initial form} of the polynomial~$f$:
\begin{equation}
   {\rm in}_\bfv(f)(\x)
   = \sum_{\bfa \in {\rm in}_\bfv(A)} c_\bfa \x^\bfa,
\end{equation}
where $A$ is the support of~$f$.
For a system~$f(\x) = \zero$ and a nonzero vector~$\bfv$,
the {\em initial form system}~${\rm in}_\bfv(f)(\x) = \zero$ is
defined by the initial forms of the polynomials~$f$
with respect to~$\bfv$.  

Because the initial coefficients of Puiseux series expansions are
solutions to initial form systems, the initial forms we consider
must have at least two monomials, otherwise the solutions will have
coordinates equal to zero and are unfit as leading coefficients in
a Puiseux series development.

\subsection{An Illustrative Example}

\begin{figure}[ht]
\begin{center}
\begin{picture}(380,200)(0,0)
\put(0,0){\epsfig{figure=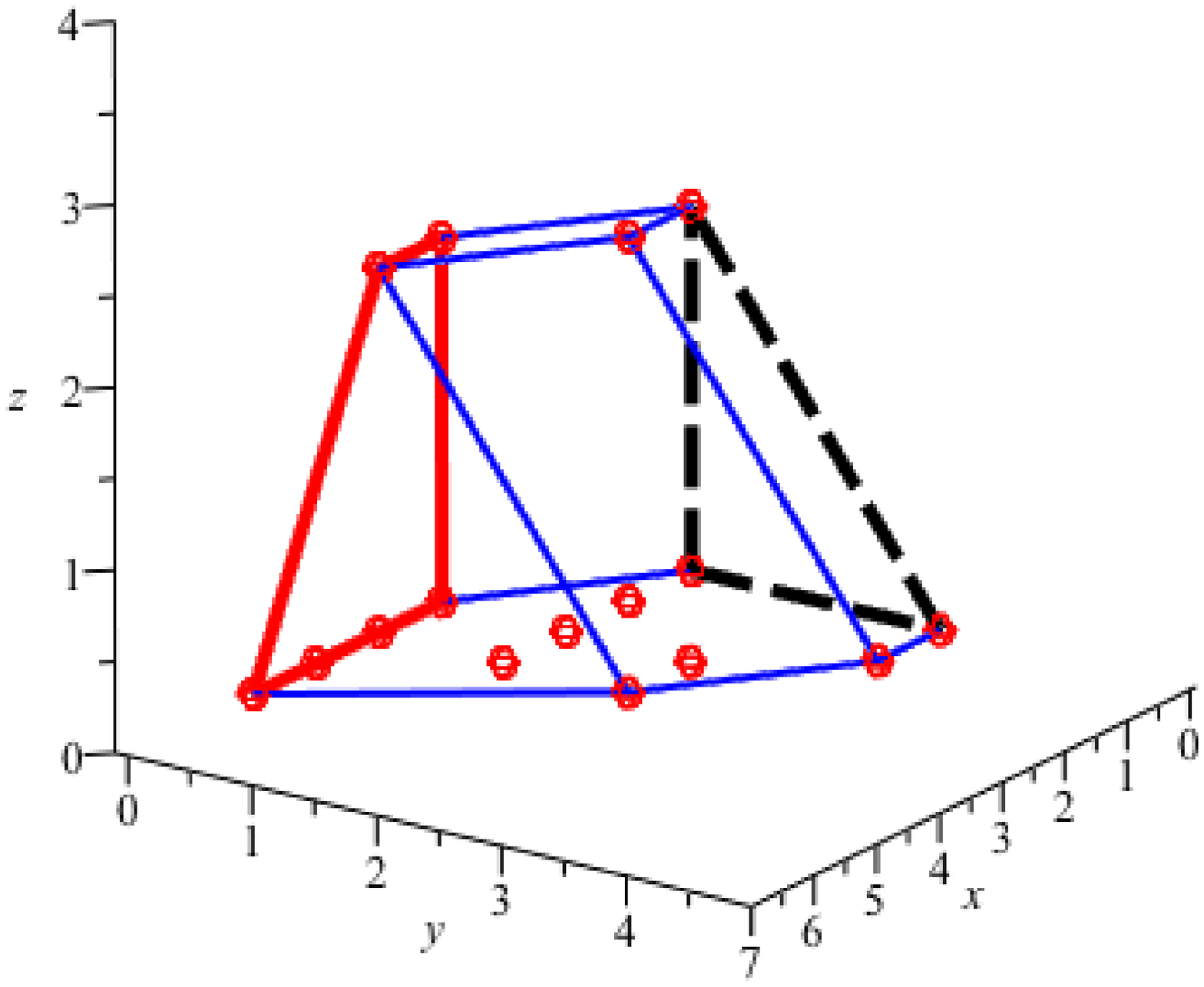, width=7cm}}
\put(190,0){\epsfig{figure=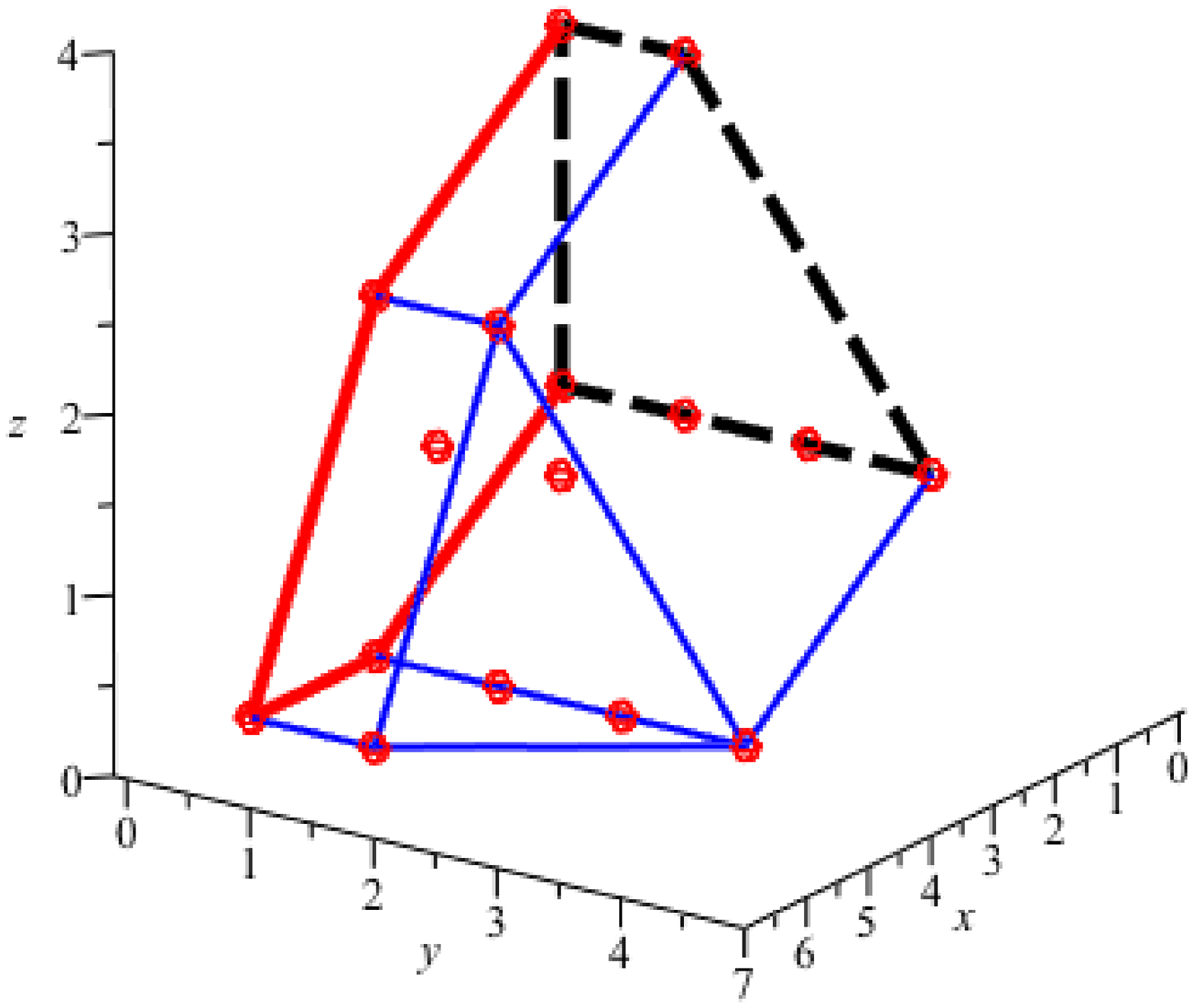, width=7cm}}
\end{picture}
\caption{From top to bottom, we see the Newton polytopes of $f_1$ and $f_2$,
  the polynomials in~(\ref{eqillusex}).
  The edges of the faces of the polytopes with normals $(1,0,0)$ 
  and~$(0,1,0)$ are marked in bold, respectively in red (thick solid lines)
  and black (thick dashed lines).}
\label{fignewtfaces1}
\end{center}
\end{figure}

\begin{figure}[hbt]
\begin{center}
\begin{picture}(180,180)(0,0)
\put(0,0){\epsfig{figure=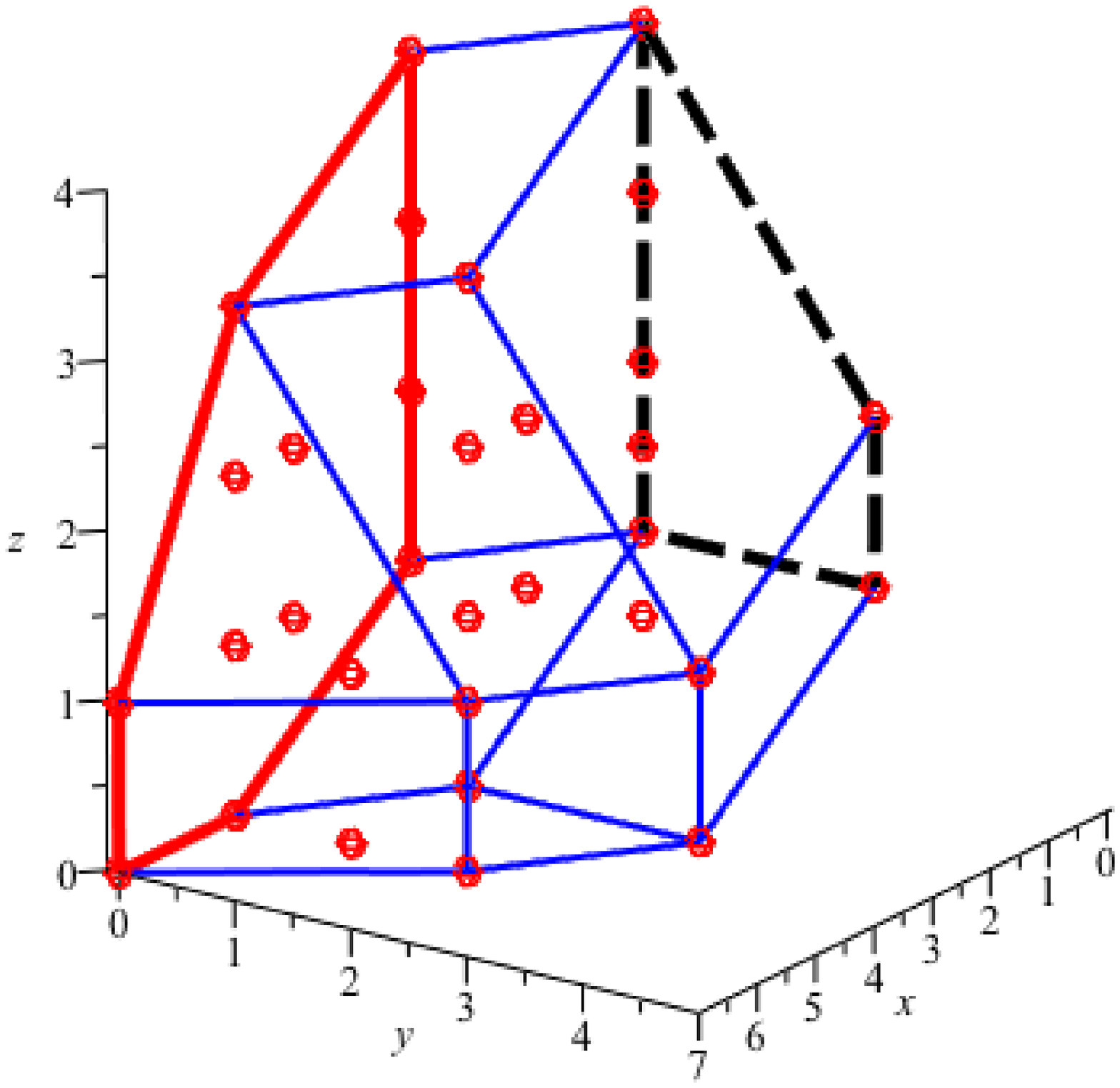, width=7cm}}
\end{picture}
\caption{The Newton polytopes of the third polynomial 
  in~(\ref{eqillusex}).
  The edges of the faces of the polytopes with normals $(1,0,0)$ 
  and~$(0,1,0)$ are marked in bold, respectively in red (thick solid lines)
  and black (thick dashed lines).}
\label{fignewtfaces2}
\end{center}
\end{figure}

In this section we indicate how the presence of a higher
dimensional solution set manifests itself from the relative
position of the Newton polytopes of the polynomials in the system.
To illustrate a numerical irreducible decomposition of the solution
set of a polynomial system, the following system was used in~\cite{SVW01a}:

\begin{equation} \label{eqillusex}
\begin{array}{l}
 f(x,y,z) = \\
 \left\{
     \begin{array}{r}
        (y - x^2 ) ( x^2 + y^2 + z^2 - 1 ) ( x - 0.5 ) = 0 \\
\vspace{-2mm} \\
        (z - x^3 ) ( x^2 + y^2 + z^2 - 1 ) ( y - 0.5 ) = 0 \\
\vspace{-2mm} \\
 (y - x^2 ) (z - x^3 ) ( x^2 + y^2 + z^2 - 1 ) ( z - 0.5 ) \! = \! 0 \\
     \end{array} \!\!
  \right.
  \end{array} 
\end{equation}
The solution set $Z = f^{-1}(\zero)$ is decomposed as
\begin{eqnarray}
Z & \!\! = \!\! & Z_2 \cup Z_1 \cup Z_0 \\
  & \!\! = \!\! & 
  \{Z_{21}\} \cup \{Z_{11} \cup Z_{12}  \cup Z_{13} \cup Z_{14} \}
   \cup \{Z_{01}\}
\end{eqnarray}
where
\begin{enumerate}
\vspace{-2mm}
\item $Z_{21}$ is the sphere $x^2 + y^2 + z^2-1=0$,
\vspace{-2mm}
\item $Z_{11}$ is the line $(x=0.5,z=0.5^3)$,
\vspace{-2mm}
\item $Z_{12}$ is the line $(x=\sqrt{0.5},y=0.5)$,
\vspace{-2mm}
\item $Z_{13}$ is the line $(x=-\sqrt{0.5},y=0.5)$,
\vspace{-2mm}
\item $Z_{14}$ is the twisted cubic $(y-x^2=0,z-x^3=0)$,
\vspace{-2mm}
\item $Z_{01}$ is the point $(x=0.5,y=0.5,z=0.5)$.
\end{enumerate}
A first cascade of homotopies in~\cite{SVW01a}
needed 197 solution paths to compute generic points on all components.
The equation-by-equation solver of~\cite{SVW08} reduced the number
of paths down to~13.  The Newton polytopes of the polynomials in 
the system are displayed 
in Figures~\ref{fignewtfaces1} and~\ref{fignewtfaces2}.

Consider a point on the 2-dimensional solution component 
of~$f^{-1}(\zero)$ and let the first coordinate of that point
go to zero.  As $x_1 = t \rightarrow 0$:
\begin{equation}
\begin{array}{l}
  {\rm in}_{(1,0,0)}(f)(x,y,z) \\
  = \left\{
     \begin{array}{r}
        y ( y^2 + z^2 - 1 ) ( - 0.5 ) = 0~\! \\
\vspace{-3mm} \\
        z ( y^2 + z^2 - 1 ) ( y - 0.5 ) = 0~\! \\
\vspace{-3mm} \\
     y z  ( y^2 + z^2 - 1 ) ( z - 0.5 ) = 0. \\
     \end{array}
  \right.
\end{array}
\end{equation}
Alternatively, as $x_2 = s \rightarrow 0$,
we end up at a solution of the initial form system:
\begin{equation}
  \begin{array}{l}
  {\rm in}_{(0,1,0)}(f)(x,y,z) \\
  = \left\{
     \begin{array}{r}
       - x^2 ( x^2 + z^2 - 1 ) ( x - 0.5 ) = 0~\! \\
\vspace{-3mm} \\
    ( z - x^3) ( x^2 + z^2 - 1 ) ( - 0.5 ) = 0~\! \\
\vspace{-3mm} \\
   - x^2 ( z - x^3) ( x^2 + z^2 - 1 ) ( z - 0.5 ) = 0. \\
     \end{array}
  \right.
\end{array}
\end{equation}

Looking at the Newton polytopes along
${\bf v} = (1,0,0)$ and ${\bf v} = (0,1,0)$,
we consider faces of the Newton polytopes, 
see Figures~\ref{fignewtfaces1} and~\ref{fignewtfaces2}.

Combining the two degenerations, we arrive at the initial form system:
\begin{equation}
\begin{array}{l}
   {\rm in}_{(0,1,0)} ({\rm in}_{(1,0,0)}(f))(x,y,z) \\
   = \left\{
      \begin{array}{r}
           y ( z^2 - 1 ) ( - 0.5 ) = 0 \\
\vspace{-3mm} \\
           z ( z^2 - 1 ) ( - 0.5 ) = 0 \\
\vspace{-3mm} \\
      y z  ( z^2 - 1 ) ( z - 0.5 ) = 0 \\
       \end{array}
    \right.
\end{array}
\end{equation}
The factor $z^2 - 1$ is shared with
${\rm in}_{(1,0,0)}({\rm in}_{(0,1,0)} (f))(x,y,z)$.

Based on these degenerations, we arrive at the following
representation for a solution surface.
The sphere is two dimensional, $x$ and $y$ are free:
\begin{equation}
  \left\{
    \begin{array}{l}
       x = t_0 \\
       y = t_1 \\
       z = 1 + c_0 t_0^2 + c_1 t_1^2.
    \end{array}
  \right.
\end{equation}
For $t_0 = 0$ and $t_1 = 0$, $z = 1$ is a solution of $z^2 - 1$ = 0.
Substituting $(x = t_0, y = t_1, z = 1 + c_0 t_0^2 + c_1 t_1^2)$
into the original system gives linear conditions on the coefficients 
of the second term: $c_0 = -0.5$ and $c_1 = -0.5$.

\subsection{Asymptotics of Algebraic Surfaces and Puiseux Series}

Denoting by~$d$ the dimension of the algebraic surface 
defined by~$f(\x) = \zero$, for $\x \in \cc^n$,
we assume the defining equations
are in Noether position so we may specialize the first~$d$
coordinates to random complex numbers in~$f(\x) = \zero$
and obtain a system with isolated solutions.
Moreover, we assume that when specializing the first~$d$ variables
to zero, the algebraic set remains of dimension~$d$.
Geometrically this means that we assume that the algebraic set
meets the first~$d$ coordinate planes (perpendicular to the
first~$d$ coordinate axes) properly.

We consider what happens when starting at a random point on the surface
we move the first $d$ coordinates to zero. 
For simplicity of notation we take $d=2$ and consider
a multiparameter family of polynomial systems:
\begin{equation} \label{eqfamily}
   \left\{
      \begin{array}{rclcl}
         f(\x) & = & \zero \\
           x_0 & = & c_0 t_0 \\
           x_1 & = & c_1 t_0^{v_{0,1}} t_1^{v_{1,1}}(c_{1,1} + O(t_0,t_1)), \\
      \end{array}
   \right.
\end{equation}
with $c_0, c_1, c_{1,1} \in \cc \setminus \{ 0 \}$,
$v_{0,1}, v_{1,1} \in \qq$, letting $t_0$ and~$t_1$ go from~1 to~0, 
starting at a generic point on the surface 
with its first two coordinates equal to~$c_0$ and~$c_1$.

The multiparameter family in~(\ref{eqfamily}) specifies
the last equation as a series to leave enough freedom for the actual
shape of the surface.  While we may always move $x_0$ as going linearly
to zero, with $x_0 = c_0 t_0$, the second coordinate of a point along
a path on the surface may no longer move linearly.  Taking $x_1$ as
$c_1 t_1$ would be too restrictive.

As we move~$x_0$ to zero as $t_0$ goes to zero, then~$x_1$ can go to
zero as well if $v_{0,1} > 0$ and $v_{1,1} > 0$,
or go to infinity if $v_{0,1} < 0$ or $v_{1,1} < 0$,
or go to $c_1 c_{1,1}$ if both $v_{0,1} = 0$ and $v_{1,1} = 0$.
The multiparameter family in~(\ref{eqfamily}) contains 
what we define as a multiparameter version of a Puiseux series
for algebraic curves.  Similar to~$x_1$, the other components of
the moving point can be developed as a generalized Puiseux series
\begin{equation} \label{eqmovsol}
   x_k = c_k t_0^{v_{0,k}} t_1^{v_{1,k}}(c_{1,k} + O(t_0,t_1)),
\end{equation}
$c_k, c_{1,k} \in \cc \setminus \{ 0 \}$, $v_{0,k}, v_{1,k} \in \qq$.
If in the limit --- when $t_0$ and $t_1$ are both zero --- 
the solution is finite and of multiplicity one,
and if the powers in the series are positive integer numbers,
then the generalized Puiseux series
coincides with a multivariate Taylor series.

As $t_0$ and $t_1$ go to zero, the system $F(t_0,t_1) = \zero$
--- obtained after replacing~$x_0$ and~$x_1$ using the last two equations
of~(\ref{eqfamily}) and after substituting~(\ref{eqmovsol}) for the
remaining $n-2$ into $f(\x) = \zero$ --- must have at least two monomials
with lowest power in~$t_0$ and lowest power in~$t_1$
in every equation because $c_k, c_{1,k} \in \cc \setminus \{ 0 \}$
for all $k = 0,1,\ldots,n-1$.  We call the part of $f(\x) = \zero$
corresponding to $F(t_0,t_1)$ with lowest powers of~$t_0$ and~$t_1$
{\em the initial form system} of $f(\x) = \zero$ with respect to the
normal vectors $\bfv_0 = (1,v_{0,1},v_{0,1},\ldots,v_{0,n-1})$
and $\bfv_1 = (0,v_{1,1},v_{1,2},\ldots,v_{1,n-1})$.
Because the normal vectors are the leading powers of the generalized
Puiseux series, $\bfv_0$ and~$\bfv_1$ can be called {\em tropisms} in
analogy to the case of algebraic curves.

The derivation of Puiseux series
for an algebraic set in any dimension~$d$ if formulated as follows.

\begin{proposition} \label{proptropisms}
If $f(\x) = \zero$ is in Noether position and defines a $d$-dimensional 
solution set in~$\cc^n$, 
intersecting the first $d$ coordinate planes in regular isolated points,
then there are $d$ linearly
independent tropisms $\bfv_0, \bfv_1, \ldots \bfv_{d-1} \in \qq^n$ so
that the initial form system 
${\rm in}_{\bfv_0} ( {\rm in}_{\bfv_1} ( 
 \cdots {\rm in}_{\bfv_{d-1}}(f) \cdots ) )(\x = \y^M) = \zero$ 
has a solution $\bfc \in (\cc \setminus \{ 0 \} )^{n-d}$.
This solution and the tropisms are the leading coefficients and powers 
of a generalized Puiseux series expansion for the algebraic set:
\begin{equation}
\begin{array}{rcl}
  x_0 & = & t_0^{v_{0,0}} \\
\\ \vspace{-5mm} \\
  x_1 & = & t_0^{v_{0,1}} t_1^{v_{1,1}} \\
      & \vdots & \\
  x_{d-1} & = & t_0^{v_{0,d-1}} t_1^{v_{1,d-1}} \cdots t_{d-1}^{v_{d-1,d-1}} \\
\\ \vspace{-5mm} \\
  x_d & = & c_0 t_0^{v_{0,d}} t_1^{v_{1,d}} \cdots t_{d-1}^{v_{d-1,d}} 
        + \cdots \\
\\ \vspace{-5mm} \\
  x_{d+1} & = & c_1 t_0^{v_{0,d+1}} t_1^{v_{1,d+1}} 
                \cdots t_{d-1}^{v_{d-1,d+1}} + \cdots \\
          & \vdots & \\
  x_n & = & c_{n-d-1} t_0^{v_{0,n-1}} t_1^{v_{1,n-1}} 
                \cdots t_{d-1}^{v_{d-1,n-1}} + \cdots \\
\end{array}
\end{equation}
\end{proposition}

\noindent {\em Proof.}  Because the set defined by $f(\x) = \zero$ is
in Noether position, we can let the first $d$ variables go to zero,
using for example a multiparameter homotopy as in~(\ref{eqfamily})
and still obtain regular isolated solutions, denoted as
$(0,0,\ldots$, $0,c_0,c_1$, $\ldots,c_{n-d-1}) \in \cc^n$.

The tropisms $\bfv_0$, $\bfv_1$, $\ldots$, $\bfv_{d-1}$
define the initial form system, i.e.: those monomials
in the system $f(\x) = \zero$ that become dominant as the parameters
$t_0$, $t_1$, $\ldots$, $t_{d-1}$ move to zero.  In particular:
for any vector~$\bfv$ in the cone spanned by the tropisms,
we have that every monomial $\x^\bfa$ in the initial form system
makes minimal inner product $\langle \bfa, \bfv \rangle$,
minimal with respect to any other monomial $\x^\bfb$ not
in the initial form system, i.e.:
$\langle \bfa, \bfv \rangle$ < $\langle \bfb, \bfv \rangle$.

Because the leading terms of the Puiseux series vanish at the 
initial form system, the inner product with the monomials and
the leading powers must be minimal compared to all other monomials
in the system.  Hence the shape of the Puiseux series. \hfill \qed

\subsection{Polyhedral Methods}

In our algorithm to develop Puiseux series developments for algebraic
sets, Proposition~\ref{proptropisms} is applied as follows.
If we are looking for an algebraic set of dimension~$d$ and 
\begin{itemize}
\item if there are no cones of vectors perpendicular to edges of the Newton
      polytopes of~$f(\x) = \zero$ of dimension~$d$, then the
      system $f(\x) = \zero$ has no solution set of dimension~$d$
      that intersects the first $d$ coordinate planes properly; otherwise
\item if a $d$-dimensional cone of vectors perpendicular to edges of the
      Newton polytopes exists, then that cone defines a part of the
      tropical prevariety.
\end{itemize}
We call a vector perpendicular to at least one edge of every
Newton polytope of~$f(\x) = \zero$ a candidate tropism or {\em pretropism}.

Algorithms to compute a tropical prevariety are described in~\cite{BJSST07}.
As we outlined in~\cite{AV11b}, we applied {\tt cddlib}~\cite{FP96}
to the Cayley embedding of the Newton polytopes of the system 
to compute pretropisms.  With the Cayley embedding we managed to compute
all pretropisms of the cyclic 12-roots problem, reported in~\cite{AV11b}.

For highly structured problems such as the cyclic $n$-roots problem,
a tropism found at lower dimension often occurs also in extended form
for higher dimensions.  For example, for $n=4$, a tropism is 
$(+1,-1,+1,-1)$ which extends directly to
$(+1,-1,+1,-1,+1,-1,+1,-1)$ for $n=8$ and
$(+1,-1,+1,-1,+1,-1,+1,-1,+1,-1,+1,-1)$ for~$n=12$,
and any $n$ that is a multiple of~4.

In addition to the extraneous results reported from the Cayley embedding,
it suffices to restrict to pretropisms with positive first coordinate
because geometrically we intersect the solution set with the coordinate
hyperplane perpendicular to the $x_0$-axes at the end of moving $x_0$
to zero.  Allowing a negative first exponent in the first pretropism
corresponds to intersecting the solution set at infinity, when in the limit
we let $x_0$ go to infinity.

In any case, after the computation of pretropisms,
exploiting permutation symmetry is relatively straightforward as we can
group the pretropisms in orbits and process only one generator per orbit.

\subsection{Puiseux Series for Algebraic Sets}

The approach to develop Puiseux series proceeds as follows.
For every $d$-dimensional cone $C$ of pretropisms:

\begin{enumerate}
\item We select $d$
      linearly independent generators to form the $d$-by-$n$ matrix~$A$
      and the corresponding unimodular transformation $\x = \y^M$.
\item Because the matrix $A$ contains pretropisms, the initial form system
      ${\rm in}_{\bfv_0} ( {\rm in}_{\bfv_1} ( 
        \cdots {\rm in}_{\bfv_{d-1}}(f) \cdots ) )(\x) = \zero$
      determined by the rows $\bfv_0$, $\bfv_1$, $\ldots$, 
      $\bfv_{d-1}$ of~$A$ has at least two monomials in every equation.
      If the initial form system has no solution with all coordinates
      different from zero, then we move to the next cone~$C$ and
      return to step 1, else we continue with the next step.
\item Solutions of the initial form system found in the previous step
      may be leading coefficients in a potential Puiseux series with
      corresponding leading powers equal to the pretropisms.
      If the leading term satisfies the entire polynomial system,
      then we report an explicit solution of the system and we
      continue processing the next cone~$C$.
      Otherwise, we take the current leading term to the next step.
\item If there is a second term in the Puiseux series,
      then we have computed an initial development for an algebraic set
      and report this development in the output.
\end{enumerate}

To compute in the last step a second term in a multivariate Puiseux
series seems very complicated, but we point out that it is not 
necessary to compute the second term in all $d$ variables.
To ensure that a solution of an initial form system is not isolated,
it suffices that we can compute a series development {\em for a curve}
starting at that solution.  In practice this means that we may
restrict all but one free variable in the series development and
apply the methods we outlined in~\cite{AV11b} for the computation
of the second term of the Puiseux series for a space curve.

With Puiseux series, the solutions of the initial form system can be
extended to form a witness set.  A witness set~\cite{SVW05} 
is a numerical data
structure for positive dimensional solution sets of polynomial systems.
Depending on the heights of the powers in the series, we may need more
than the second term to ensure convergence with Newton's method.

\section{Applications}

Our polyhedral approach enables to compute exact 
representations for positive dimensional solution sets 
of the cyclic $n$-roots problem~(\ref{eqcyclicsys}).

\subsection{On cyclic 9-roots}

Taking $n=9$ in~(\ref{eqcyclicsys}), for cyclic 9-roots, 
we show that our solution can be transformed into the same format as in 
the proof we found in~\cite[Lemma~1.1]{Fau01} of the statement
in~\cite{Bac89} that square divisors of~$n$ lead to infinitely
many cyclic $n$-roots.

Among the tropisms computed by {\tt cddlib}~\cite{FP96} on the Cayley
embedding of the Newton polytopes of the system, there is a two dimensional
cone of normal vectors spanned by
$\bfu = (1,1,-2,1,1,-2,1,1,-2)$ and $\bfv = (0,1,-1,0,1,-1,0,1,-1)$.
The vectors $\bfu$ and~$\bfv$ are tropisms.  The initial form
system ${\rm in}_\bfu ( {\rm in}_\bfv (f))(\x) = \zero$ is

\begin{equation}
 \left\{ 
   \begin{array}{rcl}
      x_{2} + x_{5} + x_{8} & = & 0 \\ 
      x_{0} x_{8} + x_{2} x_{3} + x_{5} x_{6} & = & 0 \\ 
      x_{0} x_{1} x_{2} + x_{0} x_{1} x_{8} + x_{0} x_{7} x_{8} 
      + x_{1} x_{2} x_{3} \\ 
      +~ x_{2} x_{3} x_{4} + x_{3} x_{4} x_{5} + x_{4} x_{5} x_{6} 
      + x_{5} x_{6} x_{7} \\
      +~ x_{6} x_{7} x_{8} & = & 0 \\ 
      x_{0} x_{1} x_{2} x_{8} + x_{2} x_{3} x_{4} x_{5} 
      + x_{5} x_{6} x_{7} x_{8} & = & 0 \\ 
      x_{0} x_{1} x_{2} x_{3} x_{8} + x_{0} x_{5} x_{6} x_{7} x_{8} 
      + x_{2} x_{3} x_{4} x_{5} x_{6} & = & 0 \\ 
      x_{0} x_{1} x_{2} x_{3} x_{4} x_{5} 
      + x_{0} x_{1} x_{2} x_{3} x_{4} x_{8} \\
      +~ x_{0} x_{1} x_{2} x_{3} x_{7} x_{8} 
      + x_{0} x_{1} x_{2} x_{6} x_{7} x_{8} \\
      +~ x_{0} x_{1} x_{5} x_{6} x_{7} x_{8}
      + x_{0} x_{4} x_{5} x_{6} x_{7} x_{8} \\
      +~ x_{1} x_{2} x_{3} x_{4} x_{5} x_{6} 
      + x_{2} x_{3} x_{4} x_{5} x_{6} x_{7} \\
      +~ x_{3} x_{4} x_{5} x_{6} x_{7} x_{8} & = & 0 \\ 
      x_{0} x_{1} x_{2} x_{3} x_{4} x_{5} x_{8}
      + x_{0} x_{1} x_{2} x_{5} x_{6} x_{7} x_{8} \\
      +~ x_{2} x_{3} x_{4} x_{5} x_{6} x_{7} x_{8} & = & 0 \\ 
      x_{0} x_{1} x_{2} x_{3} x_{4} x_{5} x_{6} x_{8} 
      + x_{0} x_{1} x_{2} x_{3} x_{5} x_{6} x_{7} x_{8} \\
      +~ x_{0} x_{2} x_{3} x_{4} x_{5} x_{6} x_{7} x_{8} & = & 0 \\ 
        x_{0} x_{1} x_{2} x_{3} x_{4} x_{5} x_{6} x_{7} x_{8} - 1 & = & 0.\!\!
   \end{array}
 \right.
\end{equation}
Although not binomial, ${\rm in}_\bfu ( {\rm in}_\bfv (f))(\x) = \zero$ is
is significantly sparser and thus easier to solve than the original system.
To solve ${\rm in}_\bfu ( {\rm in}_\bfv (f))(\x) = \zero$, we eliminate
$x_0$ and $x_1$ with a unimodular coordinate transformation~$M$ 
that has $\bfu$ and $\bfv$ on its first two rows.  
The last seven rows of~$M$ are zero except for the ones on the diagonal:
\begin{equation}
   M =
   \left[
   \begin{array}{rrrrrrrrr}
     1 & 1 & -2 & 1 & 1 & -2 & 1 & 1 & -2 \\
     0 & 1 & -1 & 0 & 1 & -1 & 0 & 1 & -1 \\
     0 & 0 & 1 & 0 & 0 & 0 & 0 & 0 & 0 \\
     0 & 0 & 0 & 1 & 0 & 0 & 0 & 0 & 0 \\
     0 & 0 & 0 & 0 & 1 & 0 & 0 & 0 & 0 \\
     0 & 0 & 0 & 0 & 0 & 1 & 0 & 0 & 0 \\
     0 & 0 & 0 & 0 & 0 & 0 & 1 & 0 & 0 \\
     0 & 0 & 0 & 0 & 0 & 0 & 0 & 1 & 0 \\
     0 & 0 & 0 & 0 & 0 & 0 & 0 & 0 & 1 \\
  \end{array}
  \right].
\end{equation}
The matrix~$M$ defines 
the unimodular coordinate transformation $\x = \y^M$:
\begin{equation}
   \begin{array}{l}
      x_{0} = y_{0} \\
      x_{1} = y_{0} y_{1} \\
      x_{2} = y_{0}^{-2} y_{1}^{-1} y_{2}
   \end{array}
   \begin{array}{l}
      x_{3} = y_{0} y_{3} \\
      x_{4} = y_{0} y_{1} y_{4} \\
      x_{5} = y_{0}^{-2} y_{1}^{-1} y_{5}
   \end{array}
   \begin{array}{l}
      x_{6} = y_{0} y_{6}\\
      x_{7} = y_{0} y_{1} y_{7}\\
      x_{8} = y_{0}^{-2} y_{1}^{-1} y_{8}.
   \end{array}
\end{equation}
The transformation~$\x = \y^M$ reduces the initial form system 
${\rm in}_\bfu ( {\rm in}_\bfv(f) )(\x = \y^M) = \zero$
to a system of 9 equations in 7 unknowns.

After adding two slack variables to square the system
(see~\cite{SVW05} for an illustration of introducing slack variables),
the mixed volume equals 326.  In contrast, the mixed volume of the
original polynomial system equals~20,376.

We find that the entire cyclic 9-roots system vanishes at this first term 
of the series expansion.  Recognizing the numerical roots as primitive
roots of unity leads to an exact representation of the two dimensional
set of cyclic 9-roots.

Denoting by $u = e^{i 2\pi/3}$ the primitive third root of unity,
$u^3 - 1 = 0$, our representation of the solution set is
\begin{equation} \label{eqcyc9trop}
\begin{array}{l}
   x_{0} = t_0 \\
   x_{1} = t_0 t_1 \\
   x_{2} = t_0^{-2} t_1^{-1} u^2 \\
\end{array}
\quad
\begin{array}{l}
   x_{3} = t_0 u \\
   x_{4} = t_0 t_1 u \\
   x_{5} = t_0^{-2} t_1^{-1} \\
\end{array}
\quad
\begin{array}{l}
   x_{6} = t_0 u^2 \\
   x_{7} = t_0 t_1 u^2 \\
   x_{8} = t_0^{-2} t_1^{-1} u. \\
\end{array}
\end{equation}
Introducing new variables $y_0 = t_0$, $y_1 = t_0 t_1$,
and $y_2 = t_0^{-2} t_1^{-1} u^2$, our representation becomes
\begin{equation} \label{eqcyc9equiv}
\begin{array}{l}
   x_{0} = y_0 \\
   x_{1} = y_1 \\
   x_{2} = y_2 \\
\end{array}
\quad
\begin{array}{l}
   x_{3} = y_0 u \\
   x_{4} = y_1 u \\
   x_{5} = y_2 u \\
\end{array}
\quad
\begin{array}{l}
   x_{6} = y_0 u^2 \\
   x_{7} = y_1 u^2 \\
   x_{8} = y_2 u^2 \\
\end{array}
\end{equation}
which modulo $y_0^3 y_1^3 y_2^3 u^9 - 1 = 0$ satisfies by plain
substitution the cyclic 9-roots system,
as in the proof of~\cite[Lemma~1.1]{Fau01}.

Note that the representation in~(\ref{eqcyc9trop}) allows 
a quick computation of the degree of the surface.
This degree equals the number of points in the intersection
of the surface with two random hyperplanes.
Using~(\ref{eqcyc9trop}) for points on the surface, the two
random hyperplanes become a system in the monomials
$t_0$, $t_0 t_1$, and $t_0^{-2} t_1^{-1}$:
\begin{equation}
   \left\{
      \begin{array}{c}
          \alpha_1 t_0 + \alpha_{1,2} t_0 t_1
              + \alpha_{-2,-1} t_0^{-2} t_1^{-1} = 0 \\
          \beta_1 t_0 + \beta_{1,2} t_0 t_1
              + \beta_{-2,-1} t_0^{-2} t_1^{-1} = 0 \\
      \end{array}
   \right.
\end{equation}
for some complex numbers $\alpha_{i,j}$ and $\beta_{i,j}$.
The above system is equivalent to the system
\begin{equation}
   \left\{
      \begin{array}{r}
          t_0^{-3} t_1^{-1} - c_0 = 0 \\
                        t_1 - c_1 = 0 
      \end{array}
   \right.
\end{equation}
for some $c_0, c_1 \in \cc$.  We see that for any nonzero $c_0$
and~$c_1$, the system has three solutions.
So the algebraic surface represented 
in~(\ref{eqcyc9trop}) is a cubic surface.
Using other roots of unity and permuting variables
leads to an entire orbit of cubic surfaces.

Using the representation~(\ref{eqcyc9equiv}),
we arrange the position of the coefficients with $u$
as a third root of unity:

\begin{equation} \label{eqshifting}
\begin{array}{ccc}
  1     & u     & u^{2} \\
  u     & u^{2} & 1 \\
  u^{2} & 1     & u \\
  u^{2} & u     & 1 \\
  u     & 1     & u^{2} \\
  1     & u^{2} & u \\
\end{array}
\end{equation}
shifting the variables in forward and backward order.
So the one cubic surface leads to an orbit of 6 cubic surfaces,
corresponding with our numerical results of~\cite{SVW01b}.

\subsection{On cyclic $m^2$-roots}

While the Cayley embedding becomes too wasteful to extend the
computation of {\em all} candidate tropisms beyond~$n=12$,
by the structure of the tropisms for $n=9$ 
we can predict the tropisms for cyclic 16-roots:
\begin{equation}
  \begin{array}{r}
     \bfu = (1,1,1,-3,1,1,1,-3,1,1,1,-3,1,1,1,-3), \\
     \bfv = (0,1,1,-2,0,1,1,-2,0,1,1,-2,0,1,1,-2), \\ 
     \bfw = (0,0,1,-1,0,0,1,-1,0,0,1,-1,0,0,1,-1),
  \end{array}
\end{equation}
and the corresponding initial form solutions are primitive fourth
roots of unity.  Similar to~(\ref{eqcyc9trop}) and~(\ref{eqcyc9equiv})
we can show that the exact representation obtained with tropical methods
corresponds to what is in the proof of~\cite[Lemma~1.1]{Fau01}.

A general pattern for surfaces of cyclic $m^2$-roots is below.

\begin{proposition} For $n=m^2$, there is an $(m-1)$-dimensional set
of cyclic $n$-roots, represented exactly as
\begin{equation} \label{eqm1dimset}
   \begin{array}{rcl}
       x_{km+0} & = & u_{k} t_{0} \\
       x_{km+1} & = & u_{k} t_{0} t_{1} \\
       x_{km+2} & = & u_{k} t_{0} t_{1} t_{2} \\
                & \vdots & \\
       x_{km+m-2} & = & u_{k} t_{0} t_{1} t_{2} \cdots t_{m-2} \\
       x_{km+m-1} & = & u_{k} t_{0}^{-m+1} t_{1}^{-m+2} 
                             \cdots t_{m-3}^{-2} t_{m-2}^{-1}
   \end{array}
\end{equation}
for $k=0,1,2,\ldots, m-1$ and $u_k = e^{i 2 k \pi/m}$.
\end{proposition}

The substitution $t_0 = s_0$, $t_0 t_1 = s_1$, $t_0 t_1 t_2 = s_2$,
$\ldots$, $t_{0}^{-m+1} t_{1}^{-m+2} \cdots t_{m-3}^{-2} t_{m-2}^{-1}
= s_0^{-1} s_1^{-1} \cdots s_{m-2}^{-1}$ simplifies~({\ref{eqm1dimset}).

\begin{proposition} The $(m-1)$-dimensional solution set
in~{\rm (\ref{eqm1dimset})} has degree equal to~$m$.
\end{proposition}

\noindent {\em Proof.}
To determine the degree of an $(m-1)$-dimensional algebraic set, 
we intersect the set with $m-1$ hyperplanes with random coefficients.
In any linear equation we replace the $x$-variables using the
equations in~(\ref{eqm1dimset}), dividing each equation by~$t_0$ to 
obtain a nonzero constant coefficient.  Because every $x_j$ corresponds
to one monomial in $t_0$, $t_1$, $\ldots$, $t_{m-1}$, bringing the
coefficient matrix into a reduced row echelon form leads to a
binomial system of~$m-1$ equations in~$m-1$ unknowns:

\begin{equation}
\left\{
\begin{array}{rcl}
  t_{0}^{-m}t_{1}^{-m+2}t_{2}^{-m+3} \dotsm t_{m-2}^{-1} - c_{0} & = & 0 \\
                                                   t_{1} - c_{1} & = & 0 \\
                                              t_{1}t_{2} - c_{2} & = & 0 \\
  & \vdots & \\
                        t_{1}t_{2}t_{3} \dotsm t_{m-2} - c_{m-2} & = & 0
\end{array}
\right.
\end{equation}
Collecting the coefficients $(c_0,c_1,c_2,\ldots,c_{m-2})$ in $\bf c$
and the exponents in a matrix~$A$, we denote the binomial
system as ${\bf t}^A = \bfc$ with
\begin{equation}
 A =
 \begin{bmatrix}
  -m & -m+2   & -m+3   & \cdots & -2 & -1 \\
   0  &    1   &    0   & \cdots &  0 &  0 \\
   0  &    1   &    1   & \cdots &  0 &  0 \\
 \vdots  & \vdots  &  \vdots & \ddots & \vdots & \vdots \\
   0  &    1   &    1   & \cdots &  1 &  0 \\
   0  &    1   &    1   & \cdots &  1 &  1 \\
 \end{bmatrix}.
\end{equation}

The binomial system has $|\det(A)| = m$ solutions and
therefore the degree equals~$m$.  \hfill \qed

Applying the permutation symmetry, shifting the variables forward
and backward as in~(\ref{eqshifting}),
we find $2m$ components of degree~$m$.

\bibliographystyle{plain}

\begin{thebibliography}{10}

\bibitem{AV11b}
D.~Adrovic and J.~Verschelde.
\newblock Polyhedral methods for space curves exploiting symmetry.
\newblock {\tt arXiv:1109.0241v1 [math.NA]}.

\bibitem{AV11a}
D.~Adrovic and J.~Verschelde.
\newblock Tropical algebraic geometry in {M}aple: A preprocessing algorithm for
  finding common factors to multivariate polynomials with approximate
  coefficients.
\newblock {\em Journal of Symbolic Computation}, 46(7):755--772, 2011.
\newblock Special Issue in Honour of Keith Geddes on his 60th Birthday.

\bibitem{AIL11}
F.~Aroca, G.~Ilardi, and L.~L{\'{o}}pez~de Medrano.
\newblock Puiseux power series solutions for systems of equations.
\newblock {\em International Journal of Mathematics}, 21(11):1439--1459, 2011.

\bibitem{Bac89}
J.~Backelin.
\newblock Square multiples n give infinitely many cyclic n-roots.
\newblock Reports, Matematiska Institutionen~8, Stockholms universitet, 1989.

\bibitem{BR03}
F.~Beringer and F.~Richard-Jung.
\newblock Multi-variate polynomials and {N}ewton-{P}uiseux expansions.
\newblock In F.~Winkler and U.~Langer, editors, {\em Symbolic and Numerical
  Scientific Computation Second International Conference, SNSC 2001, Hagenberg,
  Austria, September 12-14, 2001}, volume 2630 of {\em Lecture Notes in
  Computer Science}, pages 240--254, 2003.

\bibitem{Ber75}
D.N. Bernshte{\v{\i}}n.
\newblock The number of roots of a system of equations.
\newblock {\em Functional Anal. Appl.}, 9(3):183--185, 1975.
\newblock Translated from {\em Funktsional. Anal. i Prilozhen.},
  9(3):1--4,1975.

\bibitem{BJSST07}
T.~Bogart, A.N. Jensen, D.~Speyer, B.~Sturmfels, and R.R. Thomas.
\newblock Computing tropical varieties.
\newblock {\em J.\ Symbolic Computation}, 42(1):54--73, 2007.

\bibitem{Bru00}
A.D. Bruno.
\newblock {\em Power Geometry in Algebraic and Differential Equations},
  volume~57 of {\em North-Holland Mathematical Library}.
\newblock Elsevier, 2000.

\bibitem{CPHM99}
D.~Castro, L.M. Pardo, K.~H{\"{a}}gele, and J.E. Morais.
\newblock Kronecker and {N}ewton's approaches to solving: a first comparison.
\newblock {\em J.\ Complexity}, 17(1):212--303, 2001.

\bibitem{CD07}
E.~Cattani and A.~Dickenstein.
\newblock Counting solutions to binomial complete intersections.
\newblock {\em J.\ Complexity}, 23(1):82--107, 2007.

\bibitem{DMM10}
A.~Dickenstein, L.F. Matusevich, and E.~Miller.
\newblock Combinatorics of binomial primary decomposition.
\newblock {\em Mathematische Zeitschrift}, 264(4):745--763, 2010.

\bibitem{Fau01}
J.C. Faug\`ere.
\newblock Finding all the solutions of {C}yclic 9 using {G}r{\"{o}}bner basis
  techniques.
\newblock In K.~Shirayanagi and K.~Yokoyama, editors, {\em Computer Mathematics
  - Proceedings of the Fifth Asian Symposium (ASCM 2001)}, volume~9 of {\em
  Lecture Notes Series on Computing}, pages 1--12. World Scientific, 2001.

\bibitem{FP96}
K.~Fukuda and A.~Prodon.
\newblock Double description method revisited.
\newblock In M.~Deza, R.~Euler, and Y.~Manoussakis, editors, {\em Selected
  papers from the 8th Franco-Japanese and 4th Franco-Chinese Conference on
  Combinatorics and Computer Science}, volume 1120 of {\em Lecture Notes in
  Computer Science}, pages 91--111. Springer-Verlag, 1996.

\bibitem{GLS01}
M.~Giusti, G.~Lecerf, and B.~Salvy.
\newblock A {G}r{\"{o}}bner free alternative for polynomial system solving.
\newblock {\em J.\ Complexity}, 17(1):154--211, 2001.

\bibitem{GAP}
The~GAP Group.
\newblock {\em {GAP} {R}elease 4.4.12 17 {D}ecember 2008 Reference Manual},
  2008.
\newblock {\tt http://www.gap-system.org}.

\bibitem{HM06}
M.~Hampton and R.~Moeckel.
\newblock Finiteness of relative equilibria of the four-body problem.
\newblock {\em Invent.\ math.}, 163:289--312, 2006.

\bibitem{JH11}
A.~Jensen and M.~Hampton.
\newblock Finiteness of spatial central configurations in the five-body
  problem.
\newblock {\em Celestial Mechanics and Dynamical Astronomy}, 109:321--332,
  2011.

\bibitem{Jen08}
A.N. Jensen.
\newblock Computing {G}r{\"{o}}bner fans and tropical varieties in {G}fan.
\newblock In M.E. Stillman, N.~Takayama, and J.~Verschelde, editors, {\em
  Software for Algebraic Geometry}, volume 148 of {\em The IMA Volumes in
  Mathematics and its Applications}, pages 33--46. Springer-Verlag, 2008.

\bibitem{JMM07}
A.N. Jensen, H.~Markwig, and T.~Markwig.
\newblock tropical.lib. {A SINGULAR 3.0} library for computations in tropical
  geometry, 2007.
\newblock The library can be downloaded from {\tt
  http://www.mathematik.uni-kl.de/$\sim$keilen/download/Tropical/tropical.lib}.

\bibitem{JMM08}
A.N. Jensen, H.~Markwig, and T.~Markwig.
\newblock An algorithm for lifting points in a tropical variety.
\newblock {\em Collectanea Mathematica}, 59(2):129--165, 2008.

\bibitem{Kaz99}
B.~Ya. Kazarnovskii.
\newblock Truncation of systems of polynomial equations, ideals and varieties.
\newblock {\em Izvestiya: Mathematics}, 63(3):535--547, 1999.

\bibitem{McD02}
J.~McDonald.
\newblock Fractional power series solutions for systems of equations.
\newblock {\em Discrete Comput. Geom.}, 27(4):501--529, 2002.

\bibitem{Sab11}
R.~Sabeti.
\newblock Numerical-symbolic exact irreducible decomposition of cyclic-12.
\newblock {\em LMS Journal of Computation and Mathematics}, 14:155--172, 2011.

\bibitem{SVW01a}
A.J. Sommese, J.~Verschelde, and C.W. Wampler.
\newblock Numerical decomposition of the solution sets of polynomial systems
  into irreducible components.
\newblock {\em SIAM J.\ Numer.\ Anal.}, 38(6):2022--2046, 2001.

\bibitem{SVW01b}
A.J. Sommese, J.~Verschelde, and C.W. Wampler.
\newblock Using monodromy to decompose solution sets of polynomial systems into
  irreducible components.
\newblock In C.~Ciliberto, F.~Hirzebruch, R.~Miranda, and M.~Teicher, editors,
  {\em Application of Algebraic Geometry to Coding Theory, Physics and
  Computation}, pages 297--315. Kluwer Academic Publishers, 2001.
\newblock Proceedings of a NATO Conference, February 25 - March 1, 2001, Eilat,
  Israel.

\bibitem{SVW05}
A.J. Sommese, J.~Verschelde, and C.W. Wampler.
\newblock Introduction to numerical algebraic geometry.
\newblock In A.~Dickenstein and I.Z. Emiris, editors, {\em Solving Polynomial
  Equations. Foundations, Algorithms and Applications}, volume~14 of {\em
  Algorithms and Computation in Mathematics}, pages 301--337. Springer-Verlag,
  2005.

\bibitem{SVW08}
A.J. Sommese, J.~Verschelde, and C.W. Wampler.
\newblock Solving polynomial systems equation by equation.
\newblock In A.~Dickenstein, F.-O. Schreyer, and A.J. Sommese, editors, {\em
  Algorithms in Algebraic Geometry}, volume 146 of {\em The IMA Volumes in
  Mathematics and Its Applications}, pages 133--152. Springer-Verlag, 2008.

\bibitem{Sage}
W.\thinspace{}A. Stein et~al.
\newblock {\em {S}age {M}athematics {S}oftware ({V}ersion 4.5.2)}.
\newblock The Sage Development Team, 2010.
\newblock {\tt http://www.sagemath.org}.

\bibitem{Stu96}
B.~Sturmfels.
\newblock {\em Gr{\"{o}}bner {B}ases and {C}onvex {P}olytopes}, volume~8 of
  {\em University Lecture Series}.
\newblock AMS, 1996.

\bibitem{Szo11}
F.~Sz{\"{o}}ll{\H{o}}si.
\newblock {\em Construction, classification and parametrization of complex
  {H}adamard matrices}.
\newblock PhD thesis, Central European University, Budapest, 2011.
\newblock {\tt arXiv:1110.5590v1 [math.CO]}.

\bibitem{Ver99}
J.~Verschelde.
\newblock Algorithm 795: {PHC}pack: A general-purpose solver for polynomial
  systems by homotopy continuation.
\newblock {\em ACM Trans. Math. Softw.}, 25(2):251--276, 1999.
\newblock Software available at {\tt
  http://www.math.uic.edu/{\~{}}jan/download.html}.

\bibitem{Ver09b}
J.~Verschelde.
\newblock Polyhedral methods in numerical algebraic geometry.
\newblock In D.J. Bates, G.~Besana, S.~Di~Rocco, and C.W. Wampler, editors,
  {\em Interactions of Classical and Numerical Algebraic Geometry}, volume 496
  of {\em Contemporary Mathematics}, pages 243--263. AMS, 2009.

\end{thebibliography}

\end{document}